\documentclass[elsart12,epsf,eqsecnum]{revtex4}

\def\beq{\begin{equation}}

\def\eeq{\end{equation}}
\begin{document}


%

\def\ap#1#2#3{     {\it Ann. Phys. (NY) }{\bf #1} (19#2) #3}

\def\arnps#1#2#3{  {\it Ann. Rev. Nucl. Part. Sci. }{\bf #1} (19#2) #3}

\def\npb#1#2#3{    {\it Nucl. Phys. }{\bf B#1} (19#2) #3}

\def\plb#1#2#3{    {\it Phys. Lett. }{\bf B#1} (19#2) #3}

\def\prd#1#2#3{    {\it Phys. Rev. }{\bf D#1} (19#2) #3}

\def\prep#1#2#3{   {\it Phys. Rep. }{\bf #1} (19#2) #3}

\def\prl#1#2#3{    {\it Phys. Rev. Lett. }{\bf #1} (19#2) #3}

\def\ptp#1#2#3{    {\it Prog. Theor. Phys. }{\bf #1} (19#2) #3}

\def\rmp#1#2#3{    {\it Rev. Mod. Phys. }{\bf #1} (19#2) #3}

\def\zpc#1#2#3{    {\it Z. Phys. }{\bf C#1} (19#2) #3}

\def\mpla#1#2#3{   {\it Mod. Phys. Lett. }{\bf A#1} (19#2) #3}

\def\nc#1#2#3{     {\it Nuovo Cim. }{\bf #1} (19#2) #3}

\def\yf#1#2#3{     {\it Yad. Fiz. }{\bf #1} (19#2) #3}

\def\sjnp#1#2#3{   {\it Sov. J. Nucl. Phys. }{\bf #1} (19#2) #3}

\def\jetp#1#2#3{   {\it Sov. Phys. }{JETP }{\bf #1} (19#2) #3}

\def\jetpl#1#2#3{  {\it JETP Lett. }{\bf #1} (19#2) #3}


\def\ppsjnp#1#2#3{ {\it (Sov. J. Nucl. Phys. }{\bf #1} (19#2) #3}

\def\ppjetp#1#2#3{ {\it (Sov. Phys. JETP }{\bf #1} (19#2) #3}

\def\ppjetpl#1#2#3{{\it (JETP Lett. }{\bf #1} (19#2) #3} 

\def\zetf#1#2#3{   {\it Zh. ETF }{\bf #1}(19#2) #3}

\def\cmp#1#2#3{    {\it Comm. Math. Phys. }{\bf #1} (19#2) #3}

\def\cpc#1#2#3{    {\it Comp. Phys. Commun. }{\bf #1} (19#2) #3}

\def\dis#1#2{      {\it Dissertation, }{\sf #1 } 19#2}

\def\dip#1#2#3{    {\it Diplomarbeit, }{\sf #1 #2} 19#3 }

\def\ib#1#2#3{     {\it ibid. }{\bf #1} (19#2) #3}

\def\jpg#1#2#3{        {\it J. Phys}. {\bf G#1}#2#3}  

%


\voffset1.5cm


\title{Remarks on High Energy Evolution.}
\author{Alex Kovner and  Michael Lublinsky}

\address{Physics Department, University of Connecticut, 2152 Hillside
Road, Storrs, CT 06269-3046, USA}
\date{\today}

\begin{abstract}
We make several remarks on the B-JIMWLK hierarchy. First, we present a simple and instructive derivation of this equation by considering an arbitrary projectile wave function with small number of valence gluons. We also generalize the equation by including corrections which incorporate effects of high density in the projectile wave function.
Second, we systematically derive the dipole model approximation to the hierarchy. We show that in the dipole approximation the hierarchy has a simplifying property that allows its solution by solving the classical equation followed by averaging over the ensemble of initial conditions.

\end{abstract}
\maketitle
\section{Introduction.}
Recently there has been a lot of interest in the Balitsky-JIMWLK high energy evolution
\cite{balitsky,JIMWLK,cgc}. On one hand a lot of analytical and
numerical work has been done to understand the mean field version of the hierarchy due to Kovchegov \cite{kovchegov} - see for example \cite{BKT,BKN}. On the other hand recently first steps has been taken to extend this hierarchy in order to include corrections away from the large density limit which are thought to be important for generating Pomeron loops \cite{ploops},\cite{us}.

Still some analytical aspects of the hierarchy itself are not well understood. For example, the original derivation due to Balitsky is for a probe containing a fixed small number of partons. A general derivation for a small probe with an arbitrary wave function has not been given. Such a derivation would be useful not only to polish the surface, but also to help understand how to go beyond the small probe approximation. Another aspect that has been discussed in the literature \cite{nclimit,Janik} 
but not in a completely systematic way is how to get from the general JIMWLK equation to the dipole model\cite{dipole}, or large $N_c$  limit. In particular it is interesting to see whether the dipole limit allows one to probe numerically some of the fluctuations which are not included in the mean field approximation of \cite{kovchegov}, in a reasonably accessible way.

This paper contains two sketches which deal with these issues. In Section 2 we present the derivation of the JIMWLK equation for a general projectile wave function. We go beyond the standard JIMWLK expression, by including terms which arise when the gluon density in the projectile is not small. Our derivation is purely algebraic and very simple. We hope this approach will be useful for casting the whole problem in a simpler framework and thus facilitate future progress.
In Section 3 we discuss the dipole model limit of the evolution. We derive explicitly the dipole form of the evolution and also the $1/N_c$ corrections to it, starting directly from the JIMWLK equation. We make an observation that the structure of the dipole evolution is such that it allows the solution of the whole system by solving Kovchegov's equation for a single dipole $S$-matrix with subsequent averaging over the ensemble of initial conditions. We suggest some initial distributions which directly probe fluctuations in the target wave function.

\section{The S - matrix of a complex projectile.}

In this section we derive a compact and intuitive expression for the evolution of the S-matrix of a fast projectile which scatters on the fields of some hadronic target.
The equation we obtain reduces to the B-JIMWLK$^2$ equation when the projectile is small.  For arbitrary projectile we also keep certain corrections due to finite density of the projectile partons.

Although the following considerations can be applied straightforwardly to any colored particles, for simplicity in the following we consider projectiles which contain only gluons.
We will use the formalism of \cite{urs}.

Consider an arbitrary projectile whose wave function in the gluon Fock space can be written as
\begin{equation}
|\Psi\rangle=\Psi[a^{\dagger a}_i(x)]|0\rangle
\label{wf}
\end{equation}
The gluon creation operator $a^\dagger$ depends on the transverse coordinate, and also on the longitudinal momentum $k^+$. 
The projectile is assumed to have large energy. The gluon operators which enter eq.(\ref{wf}) therefore all have longitudinal momenta above some cutoff $\Lambda$. We will refer to these degrees of freedom as "valence".
Henceforth we omit the dependence on longitudinal momentum in our expressions, as the momentum enters only as a spectator variable and only determines the total phase space available for the evolution.
The wave function in eq.(\ref{wf}) is arbitrary. In particular it does not have to have a fixed number of gluon, but can be a superposition of states with different gluon numbers.

The projectile scatters on a target, which is described by some distribution of target chromoelectric fields. We denote an S-matrix of a single gluon by 
\begin{equation}
S_{ab}(x)=\langle 0| a^a_i(x)\hat Sa^{\dagger b}_i(x)|0\rangle
\label{sgluon}
\end{equation}
where $\hat S$ is the second quantized $S$-matrix operator of the field theory.
The one gluon S-matrix $S_{ab}$ does not depend on the polarization of the gluon and is diagonal in the transverse coordinate $x$. It does depend on the color fields of the target. One can think about $S_{ab}(x)$ as the eikonal $S$-matrix, although our derivation in no way depends on this. Our only assumption is that the gluons in the projectile scatter independently of each other.
Given eqs.(\ref{wf},\ref{sgluon}), the formal expression for the $S$ - matrix of the projectile can be written as
\begin{equation}
\Sigma[S]\equiv \langle\Psi|\hat S|\Psi\rangle=\langle 0|\Psi^*[a^{ a}_i(x)]
\Psi[S_{ab}(x)a^{\dagger b}_i(x)]|0\rangle
\end{equation}

As discussed at length in\cite{us}, the matrices $S(x)$ are quantum operators on the Hilbert space of the target, and thus do not commute with each. To take properly into account this noncommutativity we were led in \cite{us} to consider $S$ as a function of additional coordinate $x^-$ and to include the Wess-Zumino term in the weight function for calculating averages of functions of $S$. However for large targets this is not necessary, as the commutator of $S$ is suppressed at large density. In the present paper for simplicity we assume that the target is large, and therefore treat $S(x)$ as classical c-number fields.
The distribution of the target fields is defined by a probability density $W[S]$, so that
\begin{equation}
\int dS \,W[S]\,=\,1,\ \ \ \ \ \ \ \ \langle \Sigma\rangle\,=\,\int \,dS\,\, \Sigma[S]\,\,W[S]
\label{average}
\end{equation}
Thus the forward $S$-matrix element for the projectile $|\Psi\rangle$ on the target $W$ is given by
${\cal S}=\langle \Sigma\rangle$.

To calculate the evolution of the S-matrix we write down the evolution of the projectile wave function to first order in the change of rapidity $\delta Y$. This is given by \cite{inprep}
\begin{equation}
 |\Psi(Y+\delta Y)\rangle=\left\{\left[1-{1\over 2\pi}
\delta Y\int d^2z(b_i^a(z,[\rho])b_i^a(z),[\rho])\right]+i\int d^2z \,
b_i^a(z,[\rho])\int_{(1-\delta Y)\Lambda}^{\Lambda}{dk^+\over
\sqrt\pi |k^+|^{1/2}} a^{\dagger a}_i(k^+, z)\right\}|\Psi(Y)\rangle
 \label{wf1}
\end{equation}
Here the field $b[\rho]$ depends only on the valence degrees of freedom. It is determined as the solution of the "classical" equation of motion 
\begin{eqnarray}
&&\partial_ib_i^a(z)+gf^{abc}b^b_i(z)b^c_i(z)=g\rho^a(z)\nonumber\\
&&\epsilon_{ij}[\partial_ib^a_j-\partial_jb^a_i+gf^{abc}b^b_ib^c_j]=0
\label{b}
\end{eqnarray}
The valence charge density operator is defined as
\begin{equation}
\rho^a(z)\,\,=\,\,\int^{\infty}_{\Lambda}\,  dk^+ \,a^{\dagger b}_i(k^+,z)\,\,T^a_{bc}\,\,a^{c}_i(k^+,z)
\end{equation}
where $T^a_{bc}=if^{abc}$ is the $SU(N)$ generator in the adjoint representation. 

The expression eq.(\ref{wf1}) is not complete when the charge density is large \cite{inprep}. However it does contain all leading terms at small $\rho$ and in addition certain higher order corrections, which arise due to the fact that $b$ solves the equation eq.(\ref{b}) nonperturbatively.

The wave function of the projectile after scattering on the target field is
\begin{eqnarray}
&&\left\{\left[1-{1\over 2\pi}\delta Y\int \,d^2z\,\,(b_i^a(z,[S\rho])\,\,b_i^a(z,[S\rho]))\right]\right.+ \\
&&\,\,\,\,\,\,\,\,\quad\quad\quad\quad\,\,\,\,i\,\left.\int\, d^2z \,
b_i^a(z,[S\rho])\,\int_{(1-\delta Y)\Lambda}^{\Lambda}{dk^+\over \sqrt\pi|k^+|^{1/2}} \,
S^{ab}(z)\,\,a^{\dagger b}_i(k^+, z)\right\}\,\,
\Psi[S_{ab}(x)\,a^{\dagger b}_i(x)]\,|0\rangle \nonumber
\end{eqnarray}
Here the field $b[S\rho]$ solves the equation eq.(\ref{b}) but with the rotated source term
\begin{eqnarray}
&&\partial_ib_i^a(z,[S\rho])\,+\,g\,f^{abc}\,b^b_i(z,[S\rho])\,\,b^c_i(z,[S\rho])\,=\,g\,S^{ab}(z)\,\rho^b(z)\nonumber\\
&&\epsilon_{ij}\,
\left[\partial_ib^a_j[S\rho]\,-\,\partial_jb^a_i[S\rho]\,+\,g\,f^{abc}\,b^b_i[S\rho]\,b^c_j[S\rho]\,\right]\,=\,0
\label{sb}
\end{eqnarray}

Since the initial projectile wave function does not contain soft gluons which appear in eq.(\ref{wf1}), the calculation of the averages of the soft gluon operators is straightforward.
The S-matrix of the boosted projectile is therefore (to first order in $\delta Y$):
\begin{eqnarray}
&&\Sigma_{Y+\delta Y}[S]\,=\,\Sigma_{Y}[S]\,-\,\label{deltasigma}\\ 
&&{\delta Y\over 2\pi}\,
\langle 0|\,\Psi^*[a^{ a}_i]\,\int d^2z\,
\left[b_i^a(z,[\rho])\,b_i^a(z,[\rho])\,+\,b_i^a(z,[S\rho])\,b_i^a(z,[S\rho])\,-\,
2\,b_i^a(z,[\rho])\,S^{ba}(z)\,b_i^b(z,[S\rho])\right]\,
\Psi[S_{ab}a^{\dagger b}_i]\,|0\rangle
\nonumber
\end{eqnarray}
We now want to represent this expression in terms of functional derivatives of $\Sigma$. 
To this end we note that 
\begin{eqnarray}
\rho^a(x)\Psi[Sa^{\dagger}_i]|0\rangle&=&T^a_{bc}a^{\dagger b}_i(x){\delta\over\delta a^{\dagger c}_i(x)}\Psi[Sa^{\dagger}_i]|0\rangle=-{\rm tr} \left\{S(x)T^{a}{\delta\over \delta S^\dagger(x)}\right\}\Psi[Sa^{\dagger}_i]|0\rangle\ \ ;
\nonumber\\
\rho^b(y)\rho^a(x)\Psi[Sa^{\dagger}_i]|0\rangle&=&\left\{T^b_{cd}a^{\dagger c}_i(y){\delta\over\delta a^{\dagger d}_i(y)}\right\}\left\{T^a_{ef}a^{\dagger e}_i(x){\delta\over\delta a^{\dagger f}_i(x)}\right\}\Psi[Sa^{\dagger}_i]|0\rangle \nonumber\\
 &=&{\rm tr} \left\{S(x)T^{a}{\delta\over \delta S^\dagger(x)}\right\}{\rm tr} \left\{S(y)T^{b}{\delta\over \delta S^\dagger(y)}\right\}\Psi[Sa^{\dagger}_i]|0\rangle \ \ \ ;\nonumber\\
 \left[S(x)\rho(x)\right]^a\Psi[Sa^{\dagger}_i]|0\rangle &=&
 -{\rm tr} \left\{T^{a}S(x){\delta\over \delta S^\dagger(x)}\right\}\Psi[Sa^{\dagger}_i]|0\rangle\ \ ;\nonumber\\
 \rho^b(y) \left[S(x)\rho(x)\right]^a\Psi[Sa^{\dagger}_i]|0\rangle &=&
 {\rm tr} \left\{T^{a}S(x){\delta\over \delta S^\dagger(x)}\right\}{\rm tr} \left\{S(y)T^{b}{\delta\over \delta S^\dagger(y)}\right\}\Psi[Sa^{\dagger}_i]|0\rangle\; \ \ \ \ etc.
\label{sun}
\end{eqnarray}

The structure of these relations is quite interesting. The action of the charge density $\rho$ on the wave function is equivalent to the action of the generator of the right rotation $SU_R(N)$ on the single gluon $S$-matrix, while the action of $S\rho$ is equivalent to the action of the left rotation $SU_L(N)$. In addition the ordering of the action of the $SU(N)$ generators on $S$ is opposite to the ordering of $\rho$ acting on $a$.

Using eqs.(\ref{sun}) we can rewrite eq.(\ref{deltasigma}) as
\beq
{\delta\over\delta Y}\,\Sigma[S]\,\,=\,\,\chi\,\,\Sigma[S]
\eeq
\begin{equation}
\chi\,\,=\,\,
-\,{1\over 2\pi}\,\int\, d^2z\, \left[\tilde b^a_i(z,[J_R])\,\,\tilde b^a_i(z,[J_R])\,\,+\,\,\tilde b^a_i(z,[J_L])
\,\,\tilde b^a_i(z,[J_L])\,\,-\,\,2\,S^{ba}(z)\,\tilde b^a_i(z,[J_R])\,\,\tilde b^b_i(z,[J_L])\right]
\label{evolution}
\end{equation}
Here 
\begin{equation}
J_R^a(x)=-{\rm tr} \left\{S(x)T^{a}{\delta\over \delta S^\dagger(x)}\right\}, \ \ \ \ J_L^a(x)=
-{\rm tr} \left\{T^{a}S(x){\delta\over \delta S^\dagger(x)}\right\}, \ \ \ \  \ \ \ \ \ \ \ 
J_L^a(x)\,\,=\,\,[S(x)\,J_R(x)]^a,
\end{equation}
and the tildas over $b$ indicate that the ordering of $J$'s in $\tilde b[J]$ is opposite to the ordering 
of $\rho$ in the solution of eq.(\ref{b}) - $b[\rho]$ .
Using eq.(\ref{average}) this can be recast as the evolution equation for the target 
probability density $W$. To do this first of all we write eq.(\ref{average}) as
\begin{equation}
 \langle \Sigma\rangle_Y\,\,=\,\,\int\, dS\,\, \Sigma_{Y-Y_0}[S]\,\,W_{Y_0}[S]
\label{average1}
\end{equation}
where we think of the target as being evolved to rapidity $Y_0$, 
while the projectile takes on itself the rest of the rapidity of the process \cite{LL2}. 
The derivative of $\Sigma$ with respect to $Y$ can be replaced by the derivative with respect to $Y_0$, and the physical requirement of Lorentz invariance of the scattering amplitude allows us to differentiate $W$ rather than $\Sigma$. Subsequent integration by parts of the  functional derivatives with respect to $S$ gives the evolution equation for $W$
\beq\label{WEV}
{\delta\over\delta Y}\,W[S]\,\,=\,\,\bar\chi\,\,W[S]
\eeq
\begin{equation}
\bar\chi\,\,=\,\,-\,{1\over 2\pi}\,
\int\, d^2z\, \left[\, b^a_i(z,[J_R])\, \,b^a_i(z,[J_R])\,\,+\,\,b^a_i(z,[J_L])\,\,
 b^a_i(z,[J_L])\,\,-\,\,2\, \,b^b_i(z,[J_L])\,b^a_i(z,[J_R])\, \,S^{ba}(z)\,\right]
\label{evolutionW}
\end{equation}
where now the original ordering is restored in $b$, due to the integration by parts.

Let us take the limit of our formulae for small projectile. 
This amounts to approximating $b$ by the first order term in $g$
\begin{equation}
b^a_i(z)={g\over 2\pi}\int d^2x{(z-x)_i\over (z-x)^2}\,\rho^a(x)\,.
\end{equation}
The evolution operator $\bar\chi$ reduces to the kernel $\chi^{JIMWLK}$ 
\begin{eqnarray}
\chi^{JIMWLK}&=&-\,\frac{\alpha_s}{2\,\pi^2}
\int_{x,y,z}{(z-x)_i(z-y)_i\over (z-x)^2(z-y)^2} \,
\left[ {\rm tr} \left\{S(x)T^{a}{\delta\over 
\delta S^\dagger(x)}\right\}{\rm tr} \left\{S(y)T^{a}{\delta\over \delta S^\dagger(y)}\right\} \right.  \\
&+&\left.{\rm tr} \left\{T^{a}S(x){\delta\over \delta S^\dagger(x)}\right\}{\rm tr} 
\left\{T^{a}S(y){\delta\over \delta S^\dagger(y)}\right\}\,-\,2\,{\rm tr} \left\{T^{b}S(y){\delta\over \delta S^\dagger(y)}\right\}{\rm tr} \left\{S(x)T^{a}{\delta\over 
\delta S^\dagger(x)}\right\}\,S^{ba}(z)\right]\nonumber
\label{smallevolution}
\end{eqnarray}
In this approximation the evolution equation (\ref{WEV}) for the probability distribution $W$ becomes 
the JIMWLK equation
\begin{eqnarray}\label{smallevolutionW}
{\delta\over\delta Y}\,W[S]\,\,=\,\,\chi^{JIMWLK}\,\,W[S]\,.
\end{eqnarray}
To see that this is indeed the JIMWLK equation we recall that a simple form of the latter is written in 
terms of functional derivatives with respect to $\alpha(x,Y)$\cite{mueller} defined as
\begin{equation}
S(x)=P\exp\left\{i\int_{-\infty}^{Y} dx^- \alpha^a(x,x^-)T^a\right\}
\label{alpha}
\end{equation}
with $P$ denoting the path ordering. When acting on any function of $S$, the derivative with respect 
to $\alpha(x,Y)$ is identical to the action of $J_R(x)$:
\begin{equation}
{\delta\over\delta\alpha^e(x,Y)}F[S]={\delta S^{ab}(x)\over\delta\alpha^e(x,Y)}
{\delta\over\delta S^{ab}(x)}F[S]=i[S(x)T^e]^{ab}{\delta\over\delta S^{ab}(x)}F[S]=-iJ_R^e(x)F[S]
\label{equiv}
\end{equation}
Thus we can write eq.(\ref{smallevolutionW}) in the familiar form
\begin{equation}
{\delta\over\delta Y}W[S]=\frac{\alpha_s}{2\pi^2}\int_{x,y,z} {(z-x)_i(z-y)_i
\over (z-x)^2(z-y)^2} {\delta\over \delta \alpha^a(x,Y)}
\left[1+S^\dagger(x)S(y)-S^\dagger(x)S(z)-S^\dagger(z)S(y)\right]^{ab}
{\delta\over \delta \alpha^b(y,Y)}W[S]
\label{JIMWLK}
\end{equation}

We stress that the introduction of the additional coordinate $x^-$ and the path ordering in eq.(\ref{alpha}) serves only to mimic the action $iJ_R$ on a function of $S$ as in eq.(\ref{equiv}). In the framework of B-JIMWLK$^2$ equation it has no independent meaning, since none of the physical observables depend on $x^-$.

This completes our derivation of the B-JIMLWK$^2$ equation. We note again, that eq.(\ref{evolutionW}) 
is more general than the original B-JIMWLK$^2$. 
Expansion of the solution for the field $b$ in powers of $\rho$ furnishes corrections 
to this equation due to the high density of gluons in the projectile. 
$$\bar\chi\,\,=\,\,\chi^{JIMWLK}\,\,+\,\,O((\alpha_s\,\rho^2)^2)$$
We now turn to the discussion of the dipole model limit.

\section{The dipole model limit.}
In the large $N_c$ limit the high energy evolution can be recast in the form of the evolution of Mueller's dipoles\cite{dipole}. In this limit one thinks of the gluon as built out of a fundamental and antifundamental "constituents" $q$ and $\bar q$. The wave function then is assumed to have the dipole structure where the color charge is neutralized on pairs of points in the transverse plane, namely $\Psi[\bar q^\alpha(x)q^\alpha(y)]$.
The wave function of such a superposition of dipoles after the scattering event is
$\Psi[\bar q^\alpha(x)U^{\alpha\beta}(x,y)q^\beta(y)]$, where $U^{\alpha\beta}(x,y)=\{S_F^\dagger(x)S_F(y)\}^{\alpha\beta}$ and $S_F$ is the fundamental representation of the single gluon scattering matrix. In the large $N_c$ limit in calculating $\Sigma$ we neglect all the terms with contractions between $q$'s not belonging to the same pair, and therefore
\begin{equation}
\Sigma\,=\,\Sigma[s]\,;\,\,\,\,\,\,\,\,\,\,\,\,\,\,\,\,\,W\,=\,W[s]\,,
\end{equation}
where $s(x,y)=\frac{1}{N_c}\,{\rm Tr}[U(x,y)]$.
Given the form of the B-JIMWLK$^2$ equation we can derive the evolution of the function $\Sigma$, or alternatively $W$ and we can also see how the dipole evolution breaks down due to $1/N_c$ corrections. In line with the dipole model ideology we  
can  write down the action of the left and right rotation generators as
\begin{eqnarray}\label{dipole}
&&J_R^a(y)\Sigma[s]\,=\,\frac{1}{N_c}\,\,{\rm tr}[S^\dagger(u)S(v)\tau^a]\,\,\,
[\delta(v-y)-\delta(u-y)]\,\,{\delta \Sigma\over \delta s(u,v)}\\
&&J_L^a(y)\Sigma[s]\,=\,\frac{1}{N_c}\,{\rm tr}[S^\dagger(u)\tau^aS(v)]\,\,\,
[\delta(v-y)-\delta(u-y)]\,\,{\delta \Sigma\over \delta s(u,v)}\nonumber\\
&&J_R^a(x)J_R^a(y)\Sigma[s]\,=\,\frac{1}{N_c}\,
{\rm tr}[S^\dagger(u)S(v)\tau^a\tau^a]\,\,\,
[\delta(v-y)-\delta(u-y)]\,[\delta(v-x)-\delta(u-x)]\,\,
{\delta \Sigma\over \delta s(u,v)}\nonumber\\
&&+\,\frac{1}{N^2_c}\,{\rm tr}[S^\dagger(u)S(v)\tau^a]\,\,{\rm tr}[S^\dagger(p)S(r)\tau^a]\,\,\,
[\delta(v-y)-\delta(u-y)]\,[\delta(p-x)-\delta(r-x)]\,
{\delta^2 \Sigma\over \delta s(u,v)\delta s(p,r)}\nonumber\\
&&J_L^a(x)J_L^a(y)\Sigma[s]\,=\,\frac{1}{N_c}\,\,{\rm tr}[S^\dagger(u)S(v)\tau^a\tau^a]\,\,\,
[\delta(v-y)-\delta(u-y)]\,[\delta(v-x)-\delta(u-x)]\,\,
{\delta \Sigma\over \delta s(u,v)}\nonumber\\
&&+\,\frac{1}{N^2_c}\,{\rm tr}[S^\dagger(u)\tau^aS(v)]\,\,
{\rm tr}[S^\dagger(p)\tau^aS(r)]\,\,\,
[\delta(v-y)-\delta(u-y)]\,[\delta(p-x)-\delta(r-x)]\,
{\delta^2 \Sigma\over \delta s(u,v)\delta s(p,r)}\nonumber\\
&&J_L^b(x)J_R^a(y)\Sigma[s]
\,=\,\frac{1}{N_c}\,{\rm tr}[S^\dagger(u)\tau^bS(v)\tau^a]\,\,
[\delta(v-y)-\delta(u-y)]\,[\delta(v-x)-\delta(u-x)]\,\,{\delta \Sigma\over \delta s(u,v)}\nonumber\\
&&+\,\frac{1}{N^2_c}\,\,{\rm tr}[S^\dagger(u)S(v)\tau^a]\,\,{\rm tr}[S^\dagger(p)\tau^bS(r)]\,\,
[\delta(v-y)-\delta(u-y)]\,[\delta(p-x)-\delta(r-x)]\,\,
{\delta^2 \Sigma\over \delta s(u,v)\delta s(p,r)}\nonumber
\end{eqnarray}
Here $\tau^a$ are the generators of $SU(N)$ group in the fundamental representation. We have also dropped the subscript $F$ for brevity, but all the matrices in eq.(\ref{dipole}) are in the fundamental representation.
To simplify this further we need to use the completeness relation
\begin{equation}
\tau^a_{\alpha\beta}\tau^a_{\gamma\delta}=\frac{1}{2}
\left[\delta_{\alpha\delta}\delta_{\beta\gamma}-
{1\over N_c}\delta_{\alpha\beta}\delta_{\gamma\delta}\right ]
\end{equation}
and the identity
\beq\label{AF}
S_A^{ab}(z)\,=\,2\,Tr\,\left[ \tau^a\,S_F(z)\,\tau^b\,S_F(z)^\dagger\right]
\eeq
Working out the color algebra we find that the evolution operator $\chi^{JIMWLK}$  can be 
decomposed into two pieces 
\beq\label{dec}
\chi^{JIMWLK}[S]\,=\,\chi_{dipole}[s] \,+\,\frac{1}{N^2_c}\,\chi_{cc}[s,S]
\eeq
The dipole evolution operator
 $\chi_{dipole}$ involves dipole degrees of freedom only and it is proportional to
a single derivative with respect to $s$
\begin{eqnarray}\label{chidip}
&&\chi_{dipole}[s]\,=
\,\frac{ \bar{\alpha}_s}{2\,\pi}\,
\int_{x,y,z}\frac{(x-y)^2}{(x-z)^2\,(z\,-\,y)^2}\,\,\left[\,-s(x,\,y)\,+\,
\,s(x, z)\,s(y,z)\,\,\right]
\frac{\delta}{\delta s(x, y)} 
\end{eqnarray}
The dipole evolution operator in the form (\ref{chidip}) was first obtained in 
Ref. \cite{LL1} from a probabilistic approach to the dipole evolution (see also Refs.\cite{LL2,Janik}).
When acting on projectile`s $\Sigma$, the operators $s$ and $\frac{\delta}{\delta s}$ could be viewed 
as the dipole creation and annihilation operators respectively \cite{LL1}. On the other hand, when acting on  $W$
it is more natural to write $\chi_{dipole}$ as
\beq\label{chid}
\chi_{dipole}[s]\,=\,
\,\,\frac{ \bar{\alpha}_s}{2\,\pi}\,
\int_{x,y,z}\frac{(x-y)^2}{(x-z)^2\,(z\,-\,y)^2}\,\frac{\delta}{\delta s(x, y)}\,
\left[-\,s(x,\,y)\,+\,
\,s(x, z)\,\,s(y,z)\,\right]
\eeq
and view $s$ as the target dipole
annihilation operator while $\frac{\delta}{\delta s}$ as a target creation operator.

The operator which accounts for color correlations has the form
\beq\label{cc}
\chi_{cc}=\frac{-\bar \alpha_s}{2\pi N_c}\int_{u,v,p,r,z}K(u,v,p,r,z)
\left[Tr(S^\dagger(u)S(v)S^\dagger(r)\,S(p))\,-\,
Tr(S^\dagger(u)S(v)S^\dagger(z)\,S(r)S^\dagger(p)S(z))\right]\,
\frac{\delta^2}{\delta s(u,v)\delta s(p,r)}
\eeq
and
\beq\label{kcc}
K(u,v,r,p,z)\,=\,\left[\frac{(z-v)_i}{(z-v)^2}\,-\,\frac{(z-u)_i}{(z-u)^2}\right ]\,
\left[\frac{(z-p)_i}{(z-p)^2}\,-\,\frac{(z-r)_i}{(z-r)^2}\right]
\eeq
As is obvious from eq.(\ref{cc}), the $1/N_c$ corrections cannot be formally 
represented  in terms of dipole
degrees of freedom only and require introduction of higher order color correlations
(see also Ref. \cite{MC}). Thus strictly speaking even if at initial rapidity $W$ depends only on $s$, at rapidities of order $N_c$ this will not be the case any more. It is however unclear how important is this effect in practical terms.

Restricting ourselves to the large $N_c$ limit we obtain the following
evolution equations for the S-matrix $\Sigma$ and the target distribution function $W$ 
\begin{equation} \label{W1}
\,\frac{\partial \,\Sigma[s]}{
\partial \,Y}\,\,= \,\, \chi_{dipole}[s]\,\,\Sigma[s],
\,\,\,\,\,\,\,\,\,\,\,\,\,\,\,\,\,\,\,\,\,\,\,\,\,\,\,
\,\frac{\partial \,W(Y,[s])}{
\partial \,Y}\,\,= \,\,\chi_{dipole}[s]\,\,W(Y,[s])\,.
\end{equation}
Eq.(\ref{W1}) exhibits a curious quantum mechanical-like structure.  
The evolution in rapidity is analogous to time evolution, while the evolution operator $i\chi$
is analogous to the Hamiltonian. 
We can then think of eq.(\ref{W1}) as a Schrodinger equation for 
the wave function $W$.
The peculiarity of this quantum mechanics is that the evolution operator 
$i\chi_{dipole}$ is explicitly not Hermitian. 
This is of course because in our case the probability density is given by the 
"wave function" $W$ itself, and not by its square. The nonhermiticity of 
$\chi_{dipole}$ in the usual quantum mechanical sense is mandatory to preserve 
the total probability $\int Ds\,W[s]\,=\,1$. 
Nevertheless, also in the present case  one can sensibly talk about the 
Hilbert space of the real functionals of $s$, and define operators on 
this Hilbert space. The time evolution of these operators is given by 
the usual Heisenberg equation\cite{salam}
\begin{equation}
\label{S02}
\frac{d\,\hat O}{d\,Y}\,=\,
\left [\chi,\hat O\right ] 
\end{equation}
All that has been said so far applies equally to the dipole model and to the 
generalized
B-JIMWLK$^2$ equation discussed in the previous section. The peculiarity
of the dipole system is in the fact that the "Hamiltonian" $\chi_{dipole}$ 
is linear in the momentum conjugate to $s$. Thus the Heisenberg equation for the 
operator $\hat s$ is purely classical - the time derivative of 
$\hat s$ commutes with $\hat s$ itself:
\begin{equation}\label{S03}
\frac{d\,\hat s(x,y)}{d\,Y}\,=\,-\,\bar \alpha_s \int d^2z\,
\frac{(x-y)^2}{(x-z)^2\,(y-z)^2}\,\,\left[\hat s(x,y)\,-\,\hat s(x,z)\,\hat
s(y,z)\right ]
\end{equation}
Additionally, the observable we are interested is the $S$-matrix, which depends only on $s$ and not on its conjugate momentum. Thus the functional equation for $\Sigma$ can be replaced by an ordinary differential equation for $s$. The $S$ matrix at arbitrary time $Y$ has the form
\begin{equation}\label{S2}
 {\cal S}(Y) \,\,=\,\,\int D\,s\,\,
\,\,\Sigma[s(Y)]\,\,W[s].
\end{equation}
Where $s(Y)$ is the solution of the classical equation eq.(\ref{S03}) 
which satisfies the initial condition $s(Y=0)=s$.

Thus we see that there exists an economical way of solving the whole dipole hierarchy without relying on the mean field approximation of Kovchegov. One solves the classical equation for the dipole $S$-matrix $s$, calculates $\Sigma$ on this solution at arbitrary rapidity $Y$, and subsequently averages over the ensemble of initial conditions $W[s]$. 
This numerical program is feasible and is currently being pursued. Compared to the numerical solution of the full
JIMWLK equation (Ref. \cite{RW}) this procedure is much simpler and easier for realization as it is confined to the
large $N_c$ limit. Our goals would be to disentangle the role of the target correlations, which can be probed within the large $N_c$ limit from the effects due to the  $1/N_c$ corrections. 
The latter can be reliably estimated only by comparing solutions of eq.(\ref{S2}) with solutions of the
JIMWLK equation, provided identical initial conditions are used.  While our numerical results will be reported elsewhere, 
in the rest of this paper we discuss 
some reasonable choices for the initial ensemble $W[s]$.

The choice of the initial distribution $W[s]$ is of course tantamount to specifying 
the complete wave function of the target (or at least its large $N_c$ limit). As such it is a nonperturbative problem which can not be solved by our present means. Nevertheless, as a result of the evolution we expect the $S$ matrix to have fairly robust and universal structure. It therefore makes sense to try to model some broad features of the initial distribution which will be reflected in the preasymptotic behavior of $S$. 
One thing we can probe with the distribution rather than with a single initial condition, is the importance of the fluctuations of the target fields for the scattering matrix at large $Y$. There are two basic types of fluctuations one can consider. First, the fields in the hadron are correlated only over finite transverse distance. Thus the scattering amplitudes of dipoles which scatter at the distance larger than this correlation length from each other, must be uncorrelated. To model these fluctuations the initial ensemble    
must contain configurations which are impact parameter dependent. 
The other type of fluctuations is due to the fact that even at the same point in the transverse plane, the fields are not constant in time. The gluon density for example is bound to fluctuate as any other quantum mechanical variable. These fluctuations can be modeled within  ensemble which is translationally invariant configuration by configuration. The latter is much easier to handle numerically.
As a simple model of this kind consider the following form for the dipole scattering amplitude 
\begin{equation}
s(x,y)\,=\,e^{\,-\,(x\,-\,y)^2\,G}
\end{equation}
This is the McLerran-Venugopalan form of the dipole scattering amplitude \cite{mv}, where $G$ has the meaning of the gluon density. We can allow for the fluctuations in $G$ by simply integrating over $G$ with the weight of the form: 
\begin{equation}
W[G]\,=\,W_0\,e^{\,-\,{(G\,-\,G_0)^2\over\Delta}}
\end{equation}
This distribution allows the gluon density to fluctuate around some central value $G_0$
with the standard deviation $\Delta$.
The proper normalization of the distribution with this ansatz is 
\begin{equation}
\label{norm}
1\,=\,\int Ds\, W[s]\,=\,W_0\,\,\int_0^\infty
 dG 
\,e^{\,-\,{(G\,-\,G_0)^2\over\Delta}}
\end{equation}
which
determines the coefficient $W_0$ as
\begin{equation}
W_0\,=\,2/\,(\sqrt{\pi\,\Delta}\,(1\,+\,Erf[G_0/\sqrt{\Delta}])
\end{equation}

This simple initial distribution is summarized in the following expression for the $S$-matrix
\begin{equation}\label{an1}
 {\cal S}(Y) \,\,=\,\,W_0\,\,
\int D\,s\,dG\,\delta\left(s(x,y)\,-\,e^{-(x-y)^2G}\right)\,\,
e^{-{(G\,-\,G_0)^2\over\Delta}}\,\,\Sigma[s(Y)].
\end{equation}
with $s(Y)$ given by the solution of the classical equation eq.(\ref{S03}).
For simplicity one could consider a projectile consisting of a single dipole
\begin{equation}
\Sigma[s]\,=\,s(x,y)
\end{equation}
or for example the perturbative wave function of a virtual photon
\begin{equation}
\Sigma[s]\,=\,\int d^2(x-y)\,P^{\gamma^*}[(x-y)^2,\,Q^2]\,s(x,y)
\end{equation}
Here $P^{\gamma^*}$ stands for the well known probability to find a dipole of the
size $|x-y|$ in the wavefunction of the photon with virtuality $Q$.

By varying the width of the distribution $\Delta$ one can see how important are the local fluctuations of the gluon density. Obvious modifications of the $(x-y)$ dependence, like the perturbative $G\rightarrow G\ln[(x-y)^2\Lambda^2]$ or anomalous dimension $G\rightarrow G[(x-y)^2\Lambda^2]^{\gamma}$ are also straightforward to implement.
This distribution is translationally symmetric configuration by configuration, and thus does not take into account the finiteness of the correlation length $\lambda$ of the target fields. One can however think of it as a valid model for small size dipoles, such that $|x-y|<\lambda$.

A more realistic distribution that depends on the impact parameter can also be constructed.
To do this we must consider both $G$ and $G_0$ as $b=(x+y)/2$ dependent. 
For heavy nuclei, the impact  parameter dependence of $G_0$ can be considered as 
given by the Wood-Saxon  parameterization.
We should also allow the variance of $G$  to be uncorrelated at distances 
larger than some scale $\lambda$. In other words we would like to have
\begin{equation}
\langle \Delta(b)\Delta(0)\rangle\,=\,\Delta_0^2 e^{-b/\lambda}
\label{corr}
\end{equation}
this can be achieved simply by distributing $\Delta$ with a Gaussain measure
\begin{equation}
\exp\left\{-{1\over 2}
\int_{b,b'}\Delta(b)D(b-b')\Delta(b')\right\}
\end{equation}
Where $D(x)$ is the operator inverse to the correlator eq.(\ref{corr}). The precise form of $D$ is not particularly important, as long as its inverse decreases beyond some scale $\lambda$ and approaches a constant value $\Delta^2_0$ at zero separation.
With the appropriate normalization constant
\begin{equation}
N(b)=\int D\Delta\,W_0^{-1}(b) \,\,
\exp\left\{-{1\over 2}\int_{b,b'}
[\Delta(b)D(b-b')\Delta(b')]\right\}
\end{equation} 
the expression for the $S$-matrix becomes

\begin{eqnarray}\label{an2}
 {\cal S}(Y) &=&\int D\,s_0\,DG\,D\Delta\,\,N^{-1}[b]\,\,\times \\
& &\delta\left(s_0(x,y)-e^{-(x-y)^2G(b)}\right)\,e^{-{(G(b)\,-\,G_0(b))^2\over
\Delta(b)}}\,\,e^{\left\{-{1\over 2}\int_{b,b'}[\Delta(b)D(b-b')\Delta(b')]\right\}}
\,\,\Sigma[s(Y)]\,.\nonumber
\end{eqnarray}
The numerical implementation of this $b$-dependent measure would be very interesting, but is much less straightforward.

\acknowledgments
We are very grateful to Genya Levin for very inspiring discussions.

\end{document}